\documentclass[
prb
,twocolumn%
,amssymb, amsmath]{revtex4}
\usepackage[english]{babel}%
\usepackage{bm}%
\usepackage{graphicx}

\newcommand{\LCMO}{La$_{1-x}$Ca$_x$MnO$_3$}
\newcommand{\LSMO}{La$_{1-x}$Sr$_x$MnO$_3$}
\newcommand{\NSMO}{Nd$_{1-x}$Sr$_x$MnO$_3$}
\newcommand{\mean}[1]{\mathord{\left\langle #1 \right\rangle}}

\newcommand{\llangle}{\mathord{\langle\!\langle}}
\newcommand{\rrangle}{\mathord{\rangle\!\rangle}}

\newcommand{\cpksa}{\mathord{c^+_{\mathbf k\sigma\alpha}}}
\newcommand{\cksa}{\mathord{c_{\mathbf k\sigma\alpha}}}

\newcommand{\cbpksa}{\mathord{\bar c^+_{\mathbf k\sigma\alpha}}}
\newcommand{\cbksa}{\mathord{\bar c_{\mathbf k\sigma\alpha}}}
\newcommand{\halb}{\mathord{\frac{1}{2}}}
\newcommand{\dn}{\langle\Delta n\rangle}
\newcommand{\sksa}{\mathord{\sum_{\mathbf k\sigma\alpha}}}
\newcommand{\fk}{\mathord{\mathbf k}}
\newcommand{\fq}{\mathord{\mathbf q}}
\newcommand{\fR}{\mathord{\mathbf R}}
\newcommand{\fS}{\mathord{\mathbf S}}
\newcommand{\fs}{\mathord{\boldsymbol \sigma}}
\newcommand{\ek}{\mathord{\epsilon(\mathbf k)}}

\newcommand{\fkq}{\mathord{\mathbf k + \mathbf q}}
\newcommand{\eka}{\mathord{\epsilon_{\alpha}(\mathbf k)}}

\newcommand{\intmax}{\mathord{\int\limits^{+\infty}_{-\infty}}}

\begin{document}

\title{Extensions to the Kondo lattice model to achieve realistic Curie temperatures and appropriate behavior of the resistivity for manganites}

\author{M. Stier}
\email{stier@physik.hu-berlin.de}
\author{W. Nolting}%
\affiliation{Festk\"orpertheorie, Institut f\"ur Physik, Humboldt-Universit\"at, 12489 Berlin, Germany}
\date{\today}%

\begin{abstract}
We investigate the influence of the Jahn-Teller distortion and a direct antiferromagnetic moment coupling as extensions to a two-band Kondo lattice model for the magnetic and electronic properties of manganites. Those are calculated self-consistently via an interpolating self-energy model and a modified RKKY technique using finite Hund coupling and quantum spins. We found that both effects are essential to achieve realistic Curie temperatures if we regard intraband Coulomb repulsion. Using reliable model parameters we got results which are in very good agreement with experimental data in the whole ferromagnetic doping range. In the calculated phase diagram there are ferromagnetic metal to paramagnetic insulator transitions, accompanied by a Colossal Magnetoresistance (CMR) behavior. To improve the comparability of the measured behavior of the resistivity with the calculated one, we have to switch on interband Coulomb correlations.
\end{abstract}

\maketitle

\section{Introduction}

Manganites like \LCMO\ or \LSMO\ obtained a lot of interest in the last decades, especially after the discovery of the resistivity's dependence on the magnetic field, the colossal magnetoresistance (CMR). The detection of huge changes of the resistance in the 90's led to an acceleration of the research in this topic. Experimentalists found very rich phase diagrams containing para- and ferromagnetic insulating and metallic regions as well as different kinds of antiferromagnetism, which can be accompanied by an ordering of charge or of orbitals. Most of these phases can also be identified in theoretical calculations. But even after that long time of research there are still many unresolved issues. The origin of the CMR is determined as a competition of different phases, but the exact connections are unknown. A large variety of phases is an evidence of the complexity of those systems. Thus it is important to create a solvable model, which contains as many as possible effects that can appear in manganites.\\ \indent
The electronic and magnetic features are mainly due to the $3d$-electrons of the Mn$^{3+}$ and the Mn$^{4+}$ ions, whose ratio is defined by the doping rate $x$. Both kinds of ions have fully occupied spin-up $t_{2g}$-levels which provide a localized spin of $S=\frac{3}{2}$, but only the Mn$^{3+}$ ions also have an itinerant $e_g$-electron. The main features of this system should be covered by the Kondo lattice model (KLM) or double exchange model. But already ten years ago it became clear that this model needs to be extended to describe manganites in a proper way. First of all Millis et al. argued that the Jahn-Teller effect (JTE), which can split up the $e_g$-orbitals, should be important at low and intermediate doping range \cite{milis}. Likewise it is stated in some recent publications that a direct antiferromagnetic coupling between the localized spins, caused by a superexchange via the oxygen orbitals, is necessary to describe the competition between the different magnetic phases . Besides the necessity for the different phases, it also seems to be important to increase the CMR in theoretical calculations \cite{sen,koller,sala,alva} to be in better accordance with the measured data.

\section{Model}

In this article we want to describe the manganites in the whole ferromagnetic regime. Therefore we use a two-band KLM with finite Hund coupling. We extend this model by terms, which describe the JTE, the superexchange and the Coulomb interaction. In that general model most of the physics of the manganites should be incorporated. The phonons of the JT modes are treated classically, but the spins quantum mechanically. The model is solved approximately, but self-consistently. These extensions will appear essential to get the correct doping dependence of the Curie temperature and to achieve metal-insulator transitions simultaneously with the breakdown of the ferromagnetic order.
Therefore we study the Hamiltonian:
\begin{eqnarray}
\mathcal H &=& H_s + H_{sd} +H_{U}+ H_{AF}+H_{JT} \label{model}\\
&=& \sum_{i,j,\sigma,\alpha} T_{ij} c^+_{i\alpha\sigma}c_{j\alpha\sigma}-J_H\sum_{i,\sigma,\alpha}\boldsymbol \sigma_{i\alpha}\cdot \mathbf S_{i}+\nonumber\\
&+&\sum_{i,\sigma,\sigma' \atop \alpha,\alpha'}U^{\alpha\alpha'}_{\sigma\sigma'}n_{i\alpha\sigma}n_{i\alpha'\sigma'}+J_{AF}\sum_{\langle i,j\rangle}\mathbf S_i\cdot\mathbf S_j-\nonumber\\
&-&g\sum_{i}(Q_{2i}T^x_i+Q_{3i}T^z_i)
+\halb M\omega^2(Q^2_{2i}+Q^2_{3i})\nonumber
\end{eqnarray}
$c^{(+)}_{i\alpha\sigma}$ is the annihilator (creator) of an electron with spin $\sigma$ in the orbital $\alpha$ corresponding to the d$_{x^2-y^2}$ or d$_{3z^2-r^2}$ orbital at site $i$ and  $n_{i\alpha\sigma}=c^+_{i\alpha\sigma}c_{i\alpha\sigma}$. $J_H$ is the Hund coupling between the itinerant spin $\boldsymbol \sigma_{i\alpha}$ and the localized $\mathbf S_i$. The Hubbard parameter can represent the fact that there is only intraband Coulomb repulsion (i.e. $U^{\alpha,\alpha'\neq\alpha}_{\sigma\sigma'}=0$) or that there also is an extra interband interaction ($U^{\alpha,\alpha'\neq\alpha}_{\sigma\sigma'}\neq 0$). $J_{AF}$ corresponds to the direct superexchange. $Q_{2,3i}$ are special JT modes which couple with the pseudospin operators $T^{x,z}_i$, e.g. $T^z_i=\halb\sum_{\sigma}(n_{i\sigma\alpha=+1}-n_{i\sigma\alpha=-1})$. A similar model is used e.g. in Ref. \cite{reddy, hotta}. Possible other effects, like a hybridization between the $e_g$-bands, will not be considered explicitly. We will have a short discussion about that in the appendix \ref{app_V}.

\section{Approximation methods}
\subsection{Electronic part}

To solve the resulting equations we need some simplifications. The JT part is treated in a mean-field approximation \cite{dag}. Assuming translational invariance, introducing spherical coordinates $Q_3=Q\sin\theta,\ Q_2=Q\cos\theta$ and using "dressed" operators $\bar c^{(+)}_{i\alpha\sigma}$, which are now corresponding to the higher and lower JT orbital, yields
\begin{equation}
H_{JT}= g^2\sum_{\fk\alpha\sigma}z_{\alpha}\dn\cbpksa\cbksa\ \label{hjt}. 
\end{equation}
Here is $z_{\alpha=\pm 1}=\pm 1$ and
\begin{equation}
\dn=-\sum_{\alpha\sigma}z_{\alpha}\mean{\bar n_{\alpha\sigma}}\ .
\end{equation}
As one can see $\dn$ is dependent on the occupation numbers $\mean{\bar n_{\alpha\sigma}}$ and therefore has to be calculated self-consistently. The dressed operators are
\begin{eqnarray}
\bar c_{i\sigma\alpha=-1}&=&e^{i\theta/2}(\cos\frac{\theta}{2}c_{i\sigma,3z^2-r^2}+\sin\frac{\theta}{2}c_{i\sigma,x^2-y^2})\label{states}\\ \nonumber
\bar c_{i\sigma\alpha=+1}&=&e^{i\theta/2}(\sin\frac{\theta}{2}c_{i\sigma,3z^2-r^2}-\cos\frac{\theta}{2}c_{i\sigma,x^2-y^2})
\end{eqnarray}
and are now a superposition of the original $e_g$-states. We simplify the notation by setting $\bar c^{(+)}_{i\alpha\sigma}\rightarrow c^{(+)}_{i\alpha\sigma}$. The formulation of $H_{JT}$ in (\ref{hjt}) is possible because of the restriction to non-cooperative effects. Of course, the JT-distortions are cooperative in reality, which is e.g. important for orbital ordering. But we restrict on the non-cooperative distortions, so that we can solve the resulting equations. This suitable form of $H_{JT}$ allows us to combine it easily with the kinetic part
\begin{equation}
H_s^{JT}=\sum_{\fk\alpha\sigma}(\ek+z_{\alpha}g^2\dn)\cpksa\cksa\ ,
\end{equation}
containing the dispersion $\ek$ of the free band for simple cubic structure. In this form of $H_{JT}$ one can easily see that the JTE can split up the bands for $\dn > 0$ (band Jahn-Teller effect). That is why we will now refer to both bands as JT bands, which are denoted by the index $\alpha$. If we first leave out the antiferromagnetic coupling, we remain with a correlated KLM,
\begin{equation}
\mathcal H' = H^{JT}_s + H_{sd} + H_U\ . 
\end{equation}
The related Green's function can be approximated within an interpolating self-energy approach (ISA) \cite{isa}. In this approach we get a self-energy, which fulfills practically all limiting cases of the KLM. The exactly solvable cases of the ferromagnetically saturated semiconductor (magnetic polaron) and the atomic limit are reproducible by the ISA. Also a second order perturbation theory is included. These cases are interpolated by the use of rigorous high-energy expansion. If we postulate that this approximation is valid also between the limiting cases, we get a theory that should be reliable for all temperatures, band occupations and couplings $J_H$.\\
\indent Within the ISA, the Hubbard term was handled in an effective medium approach, so it influences the spectral weight and the width of the bands. But of course it will also give an additional splitting of the bands into different Hubbard sub-bands. We will call all bands (distinguished by the indices $(\alpha,\sigma)$ coming from the JTE and the spin) which are not affected by the Coulomb repulsion lower Hubbard bands and the other ones upper Hubbard (sub-)bands. The lower and the upper sub-bands have an energy difference of the magnitude of the Hubbard parameter $U$. For manganites the value of the Coulomb repulsion is much greater than all the other parameters so we can choose in a good approximation $U\rightarrow\infty$. This means the upper Hubbard bands will be shifted to infinite energy and will therefore never be occupied. Thus we do not have to consider these bands explicitly. We get the one particle Green's function
\begin{equation}
\llangle\cksa;\cpksa\rrangle^{\text{ISA}}_E=\hbar\frac{\gamma_{\alpha\sigma}}{E+\mu-T_{\alpha\sigma}(\fk)-\Sigma^{\text{ISA}}_{\fk\sigma\alpha}(E)}\ .\label{gf}
\end{equation}
\begin{figure}[tb]
\includegraphics[width=.95\linewidth]{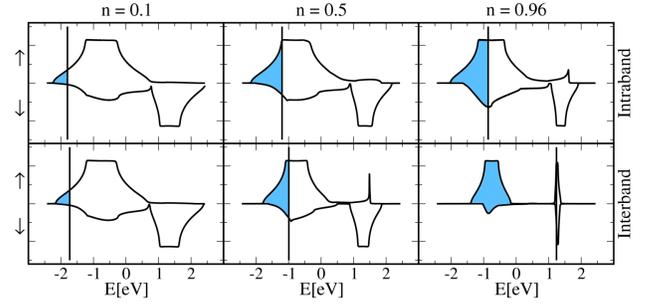}
\caption{\label{dos_mou}(Color online) Differences of the QDOS at ferromagnetic saturation ($T=0K$) if one only has intraband Coulomb interactions or additional interband repulsions. The main discrepancies are at higher occupation number $n$. Especially a complete filling of the bands is only possible with extra interband repulsion (Mott-Hubbard-insulator). The upper Hubbard bands are shifted to infinity. $J=1eV,W=3eV,g=0,J_{AF}=0$}
\end{figure}
It is 
\begin{equation}
\gamma_{\alpha\sigma} = \underbrace{1 - \mean{n_{\alpha,-\sigma}}}_{\text{intraband}}\underbrace{-\mean{n_{-\alpha,\sigma}}-\mean{n_{-\alpha,-\sigma}}}_{\text{interband}}\label{specw}
\end{equation}
the spectral weight of each band coming from the Hubbard term, $\Sigma^{\text{ISA}}_{\fk\sigma\alpha}(E)$ the self-energy in the ISA formalism and $T_{\alpha\sigma}(\fk)=z_{\alpha}g^2\dn+\gamma_{\alpha\sigma}\ek$ the centers of gravity.\\
\indent Here the $\gamma_{\alpha\sigma}$ represent the probability that there is \emph{no} repulsion partner for an electron on the same site. Thus the sum of the $\gamma_{\alpha\sigma}$, which is equivalent to the maximum occupation of the lower Hubbard bands, for the \emph{intra}band repulsion is greater than for \emph{inter}band interaction. Figure \ref{dos_mou} shows the influence of the $\gamma_{\alpha\sigma}$ on the quasi-particle density of state (QDOS). As can be seen a complete filling of the lower Hubbard bands for $n\to 1$ is only possible with the adding of interband repulsion. This complete filling correlates with the generation of a Mott-Hubbard-insulator. Thus this property of the manganites for $n=1$ is automatically fulfilled within the ISA including an extra interband repulsion.\\
\indent The explicit structure of the self-energy is 
\begin{eqnarray}
\Sigma^{\text{ISA}}_{\fk\alpha\sigma}(E)&=&-\halb z_{\sigma}J_H X_{\alpha,-\sigma}+\\
&+&\frac{1}{4}J_H^2\frac{a_{\alpha,-\sigma}G^{(0)}_{\alpha,-\sigma}(E-\halb z_{\sigma}J_H X_{\alpha,-\sigma})}{1-\halb J_H G^{(0)}_{\alpha,-\sigma}(E-\halb z_{\sigma}J_H X_{\alpha,-\sigma})}\nonumber\ \label{self},
\end{eqnarray}
where
\begin{eqnarray}
a_{\alpha\sigma}&=&S(S+1)-X_{\alpha\sigma}(X_{\alpha\sigma}+1)\nonumber\\
X_{\alpha\sigma}&=&\frac{\Delta_{\alpha\sigma}-z_{\sigma}\mean{S_z}}{1-n_{\alpha\sigma}}\\
\Delta_{\alpha\sigma}&=&\mean{S^{\sigma}_i c^+_{i\alpha,-\sigma}c_{\alpha\sigma}}+ z_{\sigma}\mean{S^z_i n_{i\alpha\sigma}}\nonumber\\
G^{(0)}_{\alpha\sigma}(E)&=&\frac{1}{N}\sum_{\fk}\frac{\hbar}{E+\mu-T_{\alpha\sigma}(\fk)}\nonumber
\end{eqnarray}
and $S^{\sigma}_i=S^x_i+i z_{\sigma}S^y_i$. With the spectral theorem we can calculate important terms like the correlation functions 
\begin{eqnarray}
\Delta_{\alpha\sigma}&=& -\frac{2}{\pi N J} \sum_{\fk}\intmax dE\ f_-(E)\times \nonumber\\
& &  \times[E-T_{\sigma\alpha}(\fk)]\text{Im}G_{\fk\sigma\alpha}(E-\mu)\label{d_isa}
\end{eqnarray}
the mean occupation values
\begin{equation}
\mean{n_{\alpha\sigma}} =  -\frac{1}{\pi N} \sum_{\fk}\intmax dE\ f_-(E)\text{Im}G_{\fk\sigma\alpha}(E-\mu) \label{n_isa}
\end{equation}
and the occupation difference
\begin{equation}
\dn = \sum_{\sigma} \left(\mean{n_{\alpha=+1,\sigma}}-\mean{n_{\alpha=-1,\sigma}}\right)\nonumber
\end{equation}
self-consistently. The only remaining model parameters are the spin $S$, $J_H$, the JT coupling $g$, the bandwidth $W$, the total density $n=1-x$ (corresponding to $\mu$) and the magnetization $\mean{S^z}$.\\
\indent There are some important changes in the QDOS of the ISA compared to e.g. mean-field calculations. The often used neglection of minority spins by setting $J_H\to\infty$ contradicts for example the KLM's exactly solvable atomic limit. Even for large $J_H$ there is a spectral weight of the spin-down peaks of the QDOS, so that a large Hund's coupling alone cannot prevent double occupation. Thus we get a finite QDOS of the spin-down electrons at lower energies (Fig. \ref{dos_mou}). This originates from scattering processes of the spin-down electrons. Those can do a spin-flip while emitting a magnon and so become a spin-up electron. This can only be done at an energy interval with a finite number of spin-up states, due to the low magnon energies. That means we have a finite occupation number of minority spin electrons even for a large Hund's coupling. Furthermore there is a peak in the spin-up spectrum at higher energies which is connected to a magnetic polaron. Details can be found in the original paper\cite{isa}.

\subsection{Magnetic part}
 
\begin{figure}[tb]
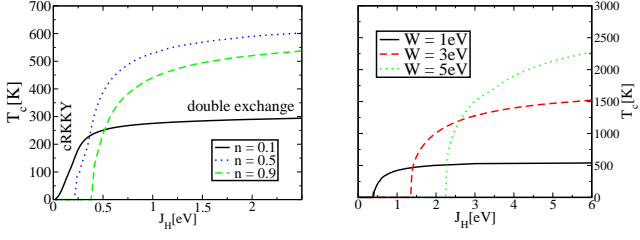

\begin{minipage}{.48\linewidth}
\includegraphics[width=.95\linewidth]{Tc_J_varn.eps}
\end{minipage}
\begin{minipage}{.03\linewidth}
\ 
\end{minipage}
\begin{minipage}{.46\linewidth}
\includegraphics[width=.95\linewidth]{Tc_J_varW2.eps}
\end{minipage}
\caption{\label{tc_j}(Color online) $T_c$-$J_H$-dependence in the mRKKY formalism. \emph{left:} For higher densities $n$ a $J_H>J_H^c$ is needed, to get a finite $T_c$. The regions where the mRKKY corresponds to conventional RKKY and the double exchange formalism are marked. $W=1eV$ \emph{right:} With increasing bandwidth $W$ a higher $T_c$ is achieved, but a larger $J_H$ is needed to get a finite Curie temperature. $n=0.95$ \emph{both:} $g=0, J_{AF}=0$, intraband repulsion}
\end{figure}

To calculate $\mean{S^z}$ in a self-consistent way, too, we use another technique. In the modified RKKY formalism (mRKKY) we try to map the Hamiltonian of the KLM onto an effective Heisenberg model \cite{mrkky1,mrkky2}. That is done by averaging out the electronic degrees of freedom. This yields
\begin{equation}
H_{ff}^{\text{eff}}=-\frac{J_H}{N}\sum_{i\sigma\sigma'\fk\fq}e^{-i\fq\fR_i}(\fS_i\cdot\boldsymbol\sigma)_{\sigma\sigma'}\mean{c^+_{\fk+\fq\sigma}c_{\fk\sigma'}}^{(s)}
\end{equation}  
$\mean{\dots}^{(s)}$ means averaging while treating the spins as numbers. We can now construct the equation of motion (EOM) for the Green's function $\hat G_{\fk,\fk+\fq}^{\sigma\sigma'}(E)=\llangle c_{\fk\sigma};c^+_{\fk+\fq\sigma'}\rrangle^{(s)}$. This yields
\begin{align}
\hat G^{\sigma\sigma'}_{\fk,\fk+\fq}& (E) =  \delta_{\fq 0}\delta_{\sigma\sigma'} G^{(0)}_{\fk}(E)-\frac{J}{2 \hbar N}\sum_{i\fk'\sigma''}\ast \\
	\ast & \Big(e^{-i(\fk-\fk')\fR_i}G^{(0)}_{\fk}(E)(\fS_i\cdot\fs)_{\sigma\sigma''} \hat G^{\sigma''\sigma'}_{\fk',\fk+\fq}(E) +\nonumber\\
	&+ e^{-i(\fk'-(\fk+\fq))\fR_i}\hat G^{\sigma''\sigma'}_{\fk,\fk'}(E)(\fS_i\cdot\fs)_{\sigma\sigma''} G^{(0)}_{\fk+\fq}(E)\Big)\nonumber\label{bew_mrkky}
\end{align}
As a simplest approximation we can replace the full Green's functions r.h.s. by free Green's functions,
\begin{eqnarray}
\hat G^{\sigma''\sigma'}_{\fk',\fk+\fq}(E) &\to &  G^{(0)}_{\fk+\fq}(E)\delta_{\fk',\fkq}\delta_{\sigma',\sigma''}\nonumber\\
\hat G^{\sigma''\sigma'}_{\fk,\fk'}(E)&\to & G^{(0)}_{\fk}(E)\delta_{\fk',\fk}\delta_{\sigma,\sigma'}
\end{eqnarray}
Then we use the spectral theorem and 
\begin{equation}
\fS_i\cdot\boldsymbol{\sigma}_i=\frac{1}{2}\left(\begin{matrix}
S_i^z & S_i^- \\ 
S_i^+ & -S_i^z
\end{matrix}\right)\nonumber
\end{equation}
to get effective exchange parameters
\begin{equation}
J(\fq)^{\text{1.O.}}=-\halb J_H^2\hbar^2\sum_{\fk}\frac{f_-(\epsilon(\fk+\fq))-f_-(\epsilon(\fk))}{\epsilon(\fk+\fq)-\epsilon(\fk)}
\end{equation}
for an effective Heisenberg model
\begin{eqnarray}
H^{\text{eff}}_{ff}&=&-\sum_{ij}\hat J_{ij}\fS_i\cdot \fS_j\\
\hat J_{ij}&=&\frac{1}{N}\sum_{\fq}\hat J(\fq)e^{-i\fq(\fR_i-\fR_j)}\nonumber\ .
\end{eqnarray}
This is the result of the conventional RKKY (cRKKY) which is originally derived by perturbation theory for low $J_H$. But we can go one step further and replace the upper Green's functions by the Green's function we derived in (\ref{gf}).
 \begin{eqnarray}
\hat G^{\sigma''\sigma'}_{\fk',\fk+\fq}(E)&\to&  G^{\sigma'}_{\fk+\fq}(E)\delta_{\fk',\fkq}\delta_{\sigma',\sigma''} \nonumber\\
\hat G^{\sigma''\sigma'}_{\fk,\fk'}(E)&\to& G^{\sigma}_{\fk}(E)\delta_{\fk',\fk}\delta_{\sigma,\sigma'}
\end{eqnarray}
After the solution of the EOM, we get effective exchange integrals 
\begin{equation}
\hat J(\fq)  = \frac{J_H^2}{4}\int^{+\infty}_{-\infty}dE\ f_-(E)\frac{1}{\pi N}\sum_{\fk\alpha\sigma}\text{Im}\ A^{\alpha\sigma}_{\fk,\fkq}(E)
\end{equation}
which define the effective Heisenberg term. These exchange integrals now contain the electronic correlations due to 
\begin{eqnarray}
A^{\alpha\sigma}_{\fk,\fkq}(E) &=& G^{(0)}_{\fk\alpha\sigma}(E)G^{\text{ISA}}_{\fkq\alpha\sigma}(E)+\nonumber\\
&&+G^{(0)}_{\fkq\alpha\sigma}(E)G^{\text{ISA}}_{\fk\alpha\sigma}(E)\\
G^{(0)}_{\fk\alpha\sigma}(E)&=&\frac{\hbar}{E+\mu-T_{\alpha\sigma}(\fk)}\nonumber
\end{eqnarray}
where $G^{\text{ISA}}_{\fk\alpha\sigma}(E)$ is the Green's function (\ref{gf}). We have now included the band index $\alpha$ which we first left out for simplicity. The mRKKY covers the cRKKY for low $J_H$ ($T_c \sim J_H^2$) and the double exchange for large couplings ($T_c\approx\text{const}$)(Fig. \ref{tc_j}). 
After the mapping onto an effective Heisenberg model, it is easy to add the antiferromagnetic part by just summing both exchange integrals
\begin{equation}
J_{ij} = \hat J_{ij} + J^{AF}_{ij}\ ,
\end{equation}
where $J^{AF}_{ij}$ has only next-neighbor elements with the magnitude $J_{AF}$. It is important to recognize the differences between both couplings. $\hat J_{ij}$ comes from the electronic properties and is a long range interaction ($\approx$ 5th-30th neighbor) and $J^{AF}_{ij}$ is constant and short range (next neighbor). Thus the adding of $J^{AF}_{ij}$ is more than just a renormalization of the energy scale. We can now directly calculate the magnetization $\mean{S^z}$ with the solution by Callen \cite{callen}:
\begin{eqnarray}
\mean{S_z}&=&\hbar\frac{(1+S+\varphi)\varphi^{2S+1}+(S-\varphi)(1+\varphi)^{2S+1}}{(1+\varphi)^{2S+1}-\varphi^{2S+1}}\label{callen}\nonumber\\
\varphi &=& \frac{1}{N}\sum_{\fq}\frac{1}{e^{\beta E(\fq)} -1}\label{sz_callen}\\
E(\fq) &=& 2\hbar\mean{S^z}\left(J_0-J(\fq)\right) -g_J\mu_B B_0\ .\nonumber
\end{eqnarray}
Furthermore one can derive equation (\ref{tc}) for the direct calculation of the Curie temperature as it is shown in the appendix.\\
\indent A typical feature of the mRKKY for larger $n$ is the appearance of a critical $J_c$, which is needed to get a finite $T_c$. This $J_c$ can become very large, especially for a larger value of the free bandwidth $W$. It can even achieve the magnitude of the large $J_H$ of the manganites (Fig. \ref{tc_j}).\\
\indent The whole model is now self-consistent with respect to the electronic and magnetic properties.

\section{Results}

\begin{figure}[tb] 
\includegraphics[width=\linewidth]{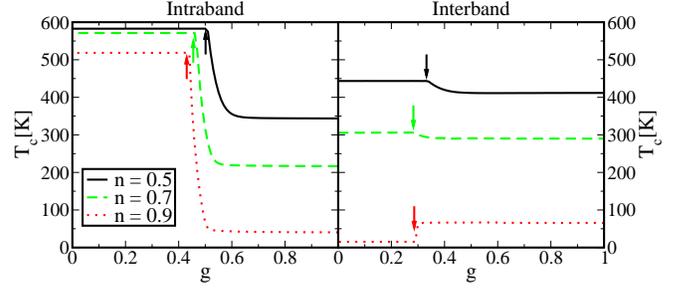}
\caption{\label{tc_g}(Color online) Influence of the JT-coupling on $T_c$. Beyond a critical $g_c$ (marked by arrows) the JT-splitting of the $e_g$-bands starts and then $T_c$ is changed. If there is only intraband Coulomb repulsion $T_c$ is everytime decreased, but can be increased if we switch on an additional interband repulsion. $J_H=2eV, W=1eV$}
\end{figure}
We will focus on the model's electronic and magnetic properties, like the resistivity, the JT-splitting and the Curie temperature. As we will see there is a great difference between having only intraband repulsion or having additional interband correlations. Interesting, for example, is the influence of the JT-splitting on the Curie temperature (Fig. \ref{tc_g}). In our self-consistent calculation we need a critical $g_c$, which was also found in other works\cite{hotta,verges,reddy}, to split up the bands . This splitting will usually decrease $T_c$, but with extra interband repulsion and larger occupation number $n$ ($n\gtrsim 0.8$, which means half-filling in this case) $T_c$ can also be increased. Thus we will observe the respective interaction in different parts.\\
\indent Experimental results show that the ferromagnetic regime is in the doping rate $0.1 \lesssim x\lesssim 0.5$ with a maximum of $T_c$ at $x\approx0.3$ \cite{phase,cheong}. The observed behavior of the Curie temperature is not reproducible in the pure two-band KLM, because it causes a finite $T_c$ for high and low doping rates. Therefore it is necessary to include other effects. In our case these are the JTE and the superexchange, incorporated by a direct antiferromagnetic exchange $J_{AF}$. Furthermore at higher electron densities a CMR is observed accompanied by a FM to PI transition. To achieve a large CMR and to reproduce the insulating behavior for $n\to 1$ we will need interband Coulomb repulsion.\\
Several suggestions were made to explain the CMR. Most are based on a competition between different phases (e.g. Ref.\cite{dag}). In Ref. \cite{sen} this is described as a result of the competing tendencies to form a charge localized or a ferromagnetic phase. This leads to a peak of the resistivity at $T_c$. A further theory is based on the current-carrier density collapse induced by the emerging of bipolarons\cite{alex1,alex2}. Those bipolarons can be broken up below $T_c$ for a exchange $J_H S$ which is large enough. We will see in our model that this peak of the resistivity at $T_c$ can be the consequence of a lattice distortion (intraband Coulomb repulsion) or of a drastic change of the spectral weight of the quasi-particle density of states (interband Coulomb repulsion).\\
\indent It is not in the scope of the paper to investigate antiferromagnetic phases or orbital ordering. Thus we will only focus on ferro- or paramagnetic phases.

\subsection{Intraband Coulomb repulsion}

First we will only use \emph{intra}band Coulomb repulsion. That means we use $\gamma_{\alpha\sigma}=1-\mean{n_{\alpha,-\sigma}}$ in (\ref{gf}).
\begin{figure}[tb] 
\includegraphics[width=\linewidth]{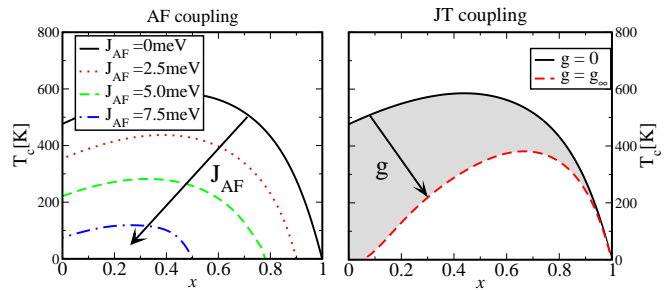}
\caption{\label{tc_n_afg}(Color online) Effect of the different couplings on the Curie temperatures at intraband Coulomb repulsion. \emph{left:} $J_{AF}$ acts on $T_c$ mainly at higher doping rates and can reduce $T_c$ to zero in this regime. This is not possible in the pure KLM. \emph{right:} The JT-coupling acts at lower doping rates and can change $T_c$ only in a special range $T^{\text{min}}(x)_c \le T_c \le T^{\text{max}}_c(x)$, in contrast to $J_{AF}$. The Curie temperature will not be reduced for $g>g_{\infty}$ any more (comp. Fig. \ref{tc_g}). The upper $T_c$-curve on each side matches the normal two-band KLM without AF and JT coupling. \emph{left:} $g=0$ \emph{right:} $J_{AF}=0$ \emph{both:} $J_H=2eV, W=1eV$}
\end{figure}
Figure \ref{tc_n_afg} shows how the JTE and the antiferromagnetic coupling act on the Curie temperature in this case. An increase of the according couplings reduces $T_c$. The impact on $T_c$ of the JTE appears at low and of the AF coupling at high doping rates. This different scope of each coupling is physically consistent, because at high electron density (that means small $x$) there are more active Mn$^{3+}$ ions, so the JTE should be supported, which is shown by measurements, too \cite{et_exp}.
A reduction of $T_c$ by the JTE has also be found by other authors, e.g. Ref.\cite{verges}. The drop of the Curie temperature at small $x$ can also be reproduced without the JTE\cite{koller} and/or if we include a large interband Hubbard interaction in our model.\\
\indent $J_{AF}$ was found to stabilize the antiferromagnetic CE-phase for $x\gtrsim0.5$ \cite{ce}, so it seems plausible that the influence of this coupling on $T_c$ should be noticed stronger in this region. The importance of $J_{AF}$ in connection with the RKKY-mechanism is even more obvious. No matter whether we use the cRKKY or the mRKKY formalism, there is always a finite Curie temperature for small electron densities, i.e. high $x$, for $J_H > 0$ (compare Ref.\cite{mrkky2} and Fig. \ref{tc_j},\ref{tc_n_afg}). Thus we need a non-RKKY-like effect to reduce $T_c$ in this region. It should favor antiferromagnetism, which is observed for higher doping rates. To the knowledge of the authors, there is no other plausible effect, which reduces $T_c$ at higher $x$ and shifts its maximum to lower doping rates and which can be added to the two-band KLM. Calculations without such an extension can achieve good results for $T_c$ at lower doping rates, but they will miss the characteristical decrease above $x\approx 0.3$.\\
\indent We can now try to calculate realistic values of $T_c$. Therefore we choose $S=\frac{3}{2}$ and fix the Hund coupling to $J_H=2eV$, which is a typical value and confirmed (with large error bars) by photoemission measurements. For the free bandwidth we choose $W=2eV$ for \LCMO\ and \NSMO, which is estimated from ab-initio calculations for \LCMO\ \cite{abin}. Since \LSMO\ is regarded to have a larger bandwidth, we use $W=3eV$. That corresponds to a hopping $t=0.16(0.25)eV$ for the simple cubic model density. The order of magnitude of the JT and AF coupling can only be determined roughly. In Ref. \cite{dag} $g$ is guessed $g=1\dots1.6\sqrt{k_{JT}t}$, which means in our model with $k_{JT}=1$ that $g_{t=0.16eV}=0.4\dots0.64\sqrt{eV}$. $J_{AF}$ is considered to be in the range of $J_{AF} =0.01\dots 0.1 t$ \cite{kapet,dag} $\Rightarrow$ $J_{AF} \approx 2\dots20meV$. These parameters will be fitted to the experiment for each material. That means we choose a set of these parameters and calculate full magnetization curves $\mean{S_z}(T)$ for each $x$ and get $T_c$ from that curves. It is necessary to calculate the full curves because of the temperature dependence of the strain and the mutual influence of the JTE and the magnetization (discussed later). If one calculates $T_c$ directly by using (\ref{tc}) one has to set $\mean{S_z}=0$ and one has no influence of the magnetization on the JT splitting any more.\\
\begin{figure}[tb] 
\includegraphics[width=\linewidth]{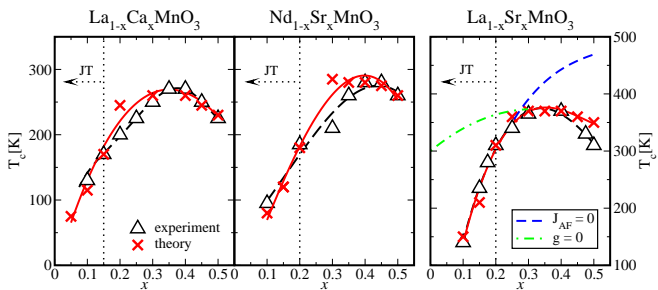}
\caption{\label{tc_n}(Color online) The calculated Curie temperatures in comparison with the experimental data for the corresponding materials in the ferromagnetic regime\cite{phase,cheong}. Lines are a guidance for the eyes. The doping range where a JT splitting occurs is marked. For \LSMO\ there are typical curves without the JTE or $J_{AF}$ but fitting the other parameters. One can see that both effects are needed to get the typical shape of the $T_c$-curves. Parameters: \LCMO: $W=2eV, J_{AF}=5.4meV,g=0.428\sqrt{eV}$ \NSMO: $W=2eV, J_{AF}=5.0meV,g=0.435\sqrt{eV}$ \LSMO: $W=3eV, J_{AF}=3.7meV,g=0.436\sqrt{eV}$ \emph{all:} $J_H=2eV$}
\end{figure}
As Fig. \ref{tc_n} shows it is possible to have an excellent agreement with the experimental data within our theory. The varied parameters $g$ and $J_{AF}$ stay in the estimated range and play a crucial role to achieve the conformity of the curves of the Curie temperatures. Firstly, which is most important for an approximative theory, our method can reproduce the right trends according to $T_c$ and secondly it can also give qualitatively right values in contrast to e.g. mean-field treatments. Thus it could be applicable to be combined with ab initio calculations to derive self-consistently input parameters like $J_H$.\\
\indent In the low doping region exists a finite JT splitting of the two JT bands of the order $E_{JT}=2g^2\dn$.
\begin{figure}[tb]
\begin{minipage}{.475\linewidth}
\begin{flushleft}
\includegraphics[width=.94\linewidth]{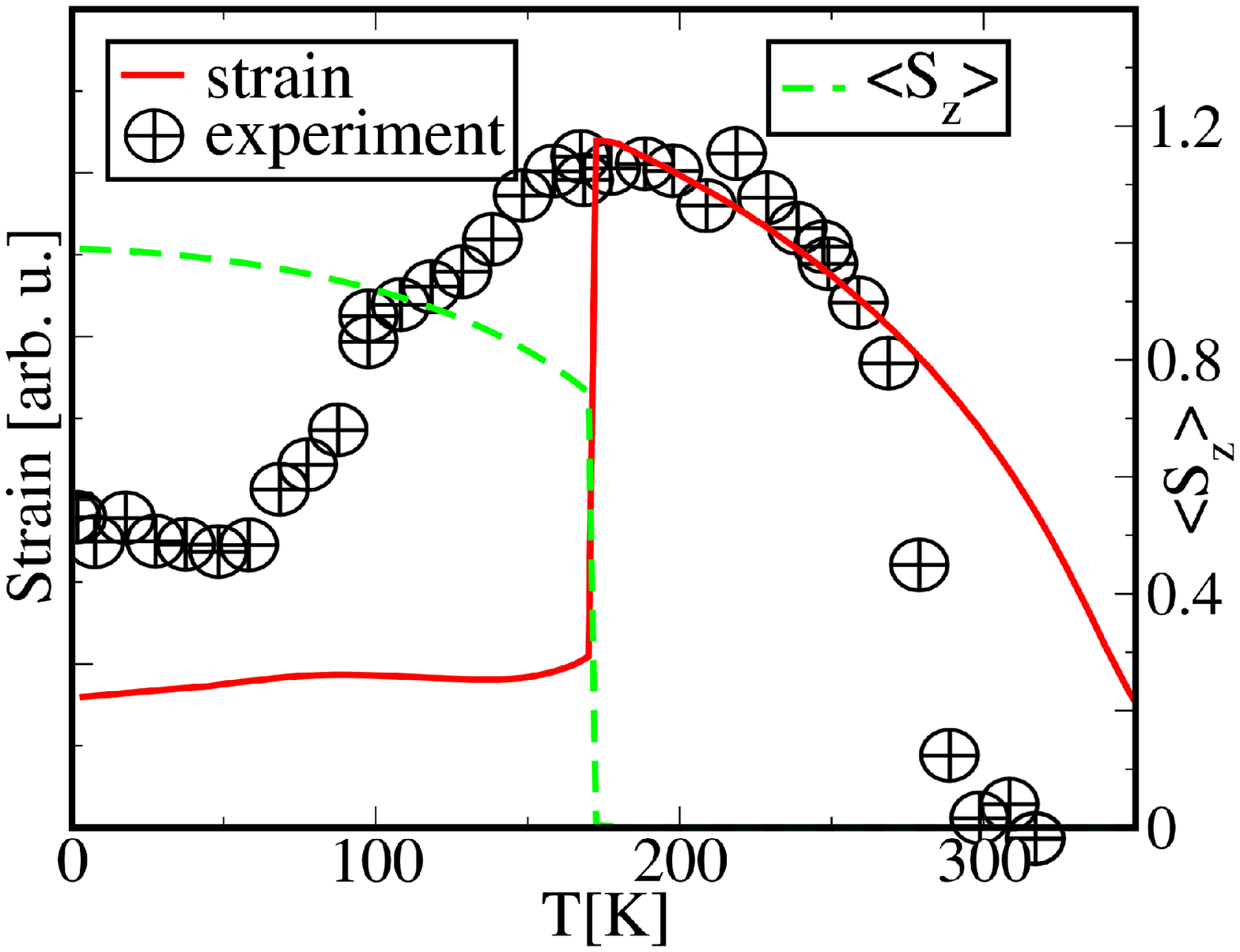}
\end{flushleft}
\end{minipage}
\begin{minipage}{.475\linewidth}
\includegraphics[width=.97\linewidth]{phase_theo2.eps}
\end{minipage}
\caption{\label{et_vgl}(Color online) \emph{left:} Comparison of the calculated values for the strain and $\mean{S_z}$ for \LCMO\ at $x=0.15$ and the measured strain \cite{et_exp}. The magnetization first lowers the JT distortion and then breaks down at the rising of the strain at $T_c$. \emph{right:} Phase diagram for the parameters of \LSMO\ in Fig. \ref{tc_n}. \emph{M} metal, \emph{I} insulator, \emph{F(P)} ferro-(para-)magnet. At the M-I transition occurs a CMR effect of a maximum of 170\% at $x=0.1$.}
\end{figure}
Near $T_c$ the occupation difference $\dn$ (equivalent to the strain $Q=g\dn$) increases, which is in qualitative agreement compared to \cite{et_exp} (Fig. \ref{et_vgl}). Even though the transition temperature of the distortion is not far away from the experimental result, the main focus should be on the qualitative behavior, because of the simple treatment of the JT-term.\\
\indent The increase of the strain is accompanied by a drastic decrease of the magnetization. That is the cause why we could not use formula (\ref{tc}) to calculate $T_c$. If we are in the critical regime of the parameter $g$ the JT splitting and the magnetization have a large impact on each other\footnote{If we are away from the critical $g_c$ (i.e. the value at which the JT splitting starts) formula (\ref{tc}) is valid and can be used to calculate $T_c$ much faster.}. At intraband repulsion both effects suppress each other. That means the larger the magnetization $\mean{S_z}$ the lower the JT splitting $E_{JT}$ and vice versa. When $\mean{S_z}$ goes down because of the rising temperature it cannot have that much effect on the JT splitting any more. Therefore the splitting of the JT bands becomes stronger. With this increase of $E_{JT}$ it will now suppress the magnetization even more and will lower it in addition to the lowering of the temperature effect. When this self-energizing effect exceeds a special value, we have a first-order transition of the magnetization. One can also see this behavior in the QDOS (Fig. \ref{dos}) where the different bands split more for $T=T_c$. The finite QDOS for the spin-down electrons at lower energies is a quantum mechanical effect due to scattering\cite{isa}.\\
\indent If we calculate the resistivity according to (\ref{sigma_el}) and define an insulator/metal via the temperature behavior, we can get the phase diagram of Fig. \ref{et_vgl}. There are metal-insulator transitions at lower doping rates, which are accompanied by a change of the magnetic phase. In Fig. \ref{rho_vgl} (left) the explicit resistivity curves are plotted. One sees the jump of $\rho_{\text{el}}$ at $T_c$ for low doping rates. The reason for that is the behavior of the JT-splitting (comp. Fig. \ref{et_vgl}, left). At $T_c$ the bands split which means a lower total DOS at the Fermi energy (comp. Fig. \ref{dos}) and therefore a decrease of the conductivity. With rising temperature the JT-splitting becomes smaller and the bands get more and more overlap again. Thus the resistivity decreases with temperature resulting in an insulating behavior. With the breakdown of the JT splitting we get the normal metallic phase again. Though the simultaneous FM/PI transition leads to a CMR behavior, the resistivity jump is too small compared to measured results.\\
\indent Antiferromagnetic phases and phase separation are very important, but were not part of this work, because it would enormously expand the complexity of the problem. Those were found in Ref.\cite{john} for the KLM using the same self-energy. The inclusion of the AF-phases within our whole model will be left for later investigations.
\begin{figure}[tb]
\includegraphics[width=\linewidth]{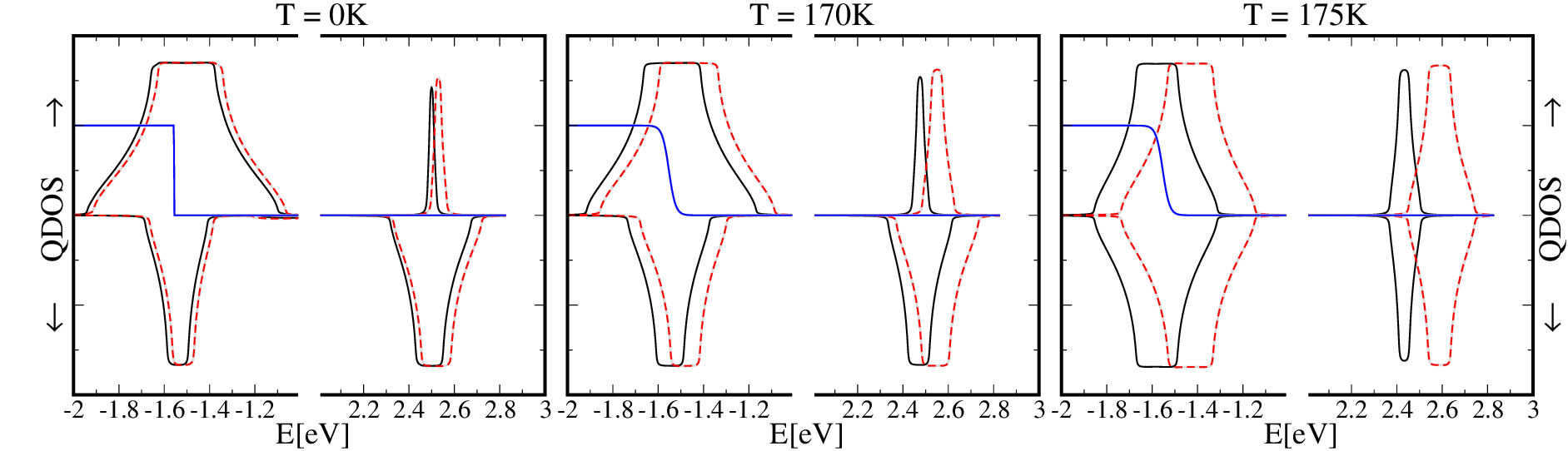}
\caption{\label{dos}(Color online) The quasi-particle DOS and fermi function (blue solid line) for the parameters of \LCMO\ in Fig. \ref{tc_n} at $x=0.15$, below and at $T_c=175K$. Those bands originate from the combined states of $e_g$-orbitals shown in (\ref{states}). One sees that the splitting $\dn$ between the lower (solid line) and the upper (dashed line) JT band rises at $T_c$ and the same occurs for the strain $Q=g\dn$ in accordance to the experiment \cite{et_exp}. For a large $J_H$ there is a finite occupation of the spin-down band for $T<T_c$, too.}
\end{figure}

\subsection{Interband Coulomb repulsion}
\begin{figure}[tb]
\includegraphics[width=.95\linewidth]{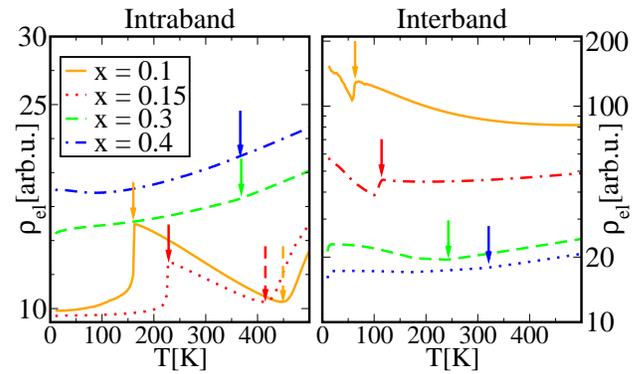}
\caption{\label{rho_vgl}(Color online) \emph{left:} Calculated resistivities for the parameters of \LSMO\ in Fig. \ref{tc_n}, intraband repulsion \emph{right:} Calculated resistivities with extra \emph{inter}band repulsion: $W=2eV, J=3eV,J_{AF}=0, g = 0.5\sqrt{eV}$ The values of $T_c$ are marked by solid arrows and the critical temperatures of the JTE by dashed ones. With extra interband repulsion the curves now show qualitative agreement with the experiment\cite{wid} and especially a ferromagnetic insulating phase can be achieved at lower doping rates. }
\end{figure}

\begin{figure}[tb]
\includegraphics[width=\linewidth]{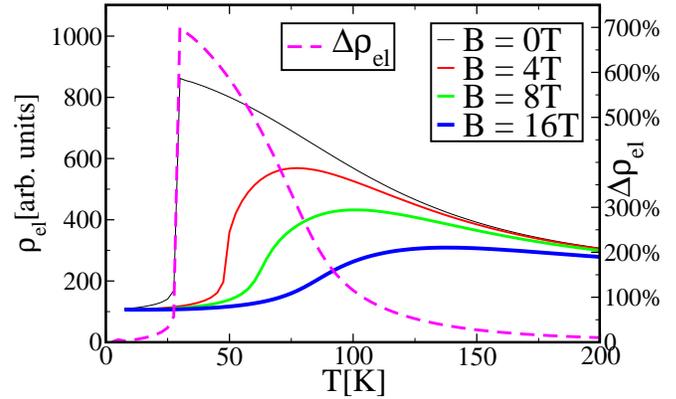}
\caption{\label{cmr1}(color online) The resistivity for different external magnetic fields. With extra interband Hubbard-repulsion CMR-behavior with $\Delta \rho = \frac{\rho(H=0)-\rho(H_{\text{max}})}{\rho(H_{\text{max}})}$ over 700\% can be found. $x=0.1, g=0\sqrt{eV},W=3eV,J=3eV$}
\end{figure}

\begin{figure}[tb]
\includegraphics[width=.95\linewidth]{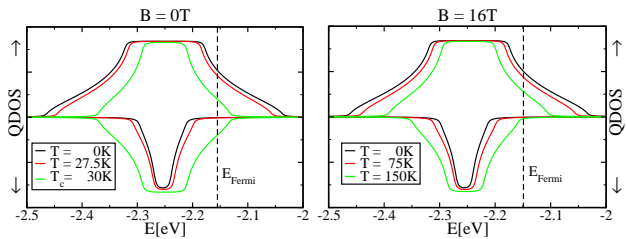}
\caption{\label{dos_c}(color online) The quasi-particle DOS (solid lines) for parameters of Fig. \ref{cmr1} at different temperatures. Only the lower part of the QDOS closed to the Fermi edge ($\approx -2.15eV$, dashed line) is shown. \emph{left:} ($B=0T$) The QDOS at the Fermi energy is changing very drastically at $T_c$. It develops a lack of spin-up electrons which cannot be compensated by the spin-down electrons and creates the jump of the resistivity in Fig. \ref{cmr1}. After $T_c$ the QDOS is not changing very much and $\rho_{\text{el}}$ lowers because of the softening of the Fermi edge. \emph{right:} ($B=16T$) The QDOS is not changing very much at all. Thus it exists just a small change of $\rho_{\text{el}}$.}
\end{figure}

With intraband repulsion we achieved very good results concerning the  Curie temperatures and the JT splitting. Even though there are important phases and phase transitions (FM-PI), the ferromagnetic insulating (FI) phase is missing and, of course, the many antiferromagnetic and charge/orbital ordering phases are not there, either. While we will not consider antiferromagnetism and orbital/charge ordering, we will try to get the FI phase in our model. Such a phase is found for low doping rates, e.g. in Ref.\cite{wid} . This phase cannot be reproduced in the intraband model. These measurements also show that the resistivities decrease with increasing doping rates. That is not reproduced in the intraband treatment, thus we will now add the interband Coulomb interaction.\\
\indent Actually we can achieve a ferromagnetic phase with insulating behavior of the resistivity (Fig. \ref{rho_vgl}). Additionally we have a better comparability of the theoretical and experimental results for the resistivity in the whole ferromagnetic regime, too. In particular the system becomes an insulator at $x\to 0$, the Mott-insulator as described in Fig \ref{dos_mou}. But the ferromagnetism breaks down in this limit ($T_c=0K$) which indicates that other phases are important for $x\to 0$. This is indeed the case, as it is widely known from the experiments showing antiferromagnetic or spin glass phases, which we are not treating in this work. That our approximative self-energy $\Sigma^{\text{ISA}}_{\sigma\alpha}(E)$ actually leads to antiferromagnetism at $x\to 0$ was shown by Hennig\cite{john}.\\
\indent Furthermore we can get a high CMR effect (Fig. \ref{cmr1}) which was very much smaller while using only intraband repulsion. In this case the origin of the resistivity jump is different from that which occurred in the intraband calculations. The latter originated from the JTE and the increasing band splitting. Now, in the interband calculations, it comes from a drastic change of the DOS in the bands themselves. Figure \ref{dos_c} shows that at $T_c$ it comes to a large rearranging of the DOS at the Fermi energy if no external magnetic field is applied. This leads to a drastic reduction of the spin-up DOS which cannot be compensated by the increase of the spin-down DOS. Therefore the resistivity is jumping to a higher value. At higher temperatures above $T_c$ there is an increase of the conductivity due to the softening of the Fermi function. The fast changing of the DOS can be slowed down in the presence of an external magnetic field and leads to a delay of the increase of $\rho_{\text{el}}$. That means we have a CMR effect.\\
\indent Quantitatively, the resistivity should vary over more orders of magnitude, according to the experiment. We also have to pay for the better agreement of the resistivity with a worse behavior of $T_c$. The Curie temperature is much more suppressed if we use interband repulsion and we cannot get the typical shape of a parabola for the $T_c$-$n$-curves. Such a strong influence on $T_c$ is probably due to the effective medium treatment of the Hubbard part. This treatment cannot contain all many-body correlation effects. Thus for the calculation of $T_c$ we should include the Coulomb interaction in a more subtle way, which is no simple task, of course.
\noindent
\noindent
\section{Summary}

We investigated manganite systems with a two-band KLM, which was extended by terms that represent the Coulomb correlations, the JTE and the superexchange. Within this model we calculated the electronic and magnetic properties self-consistently by the use of an interpolating self-energy approach and a modified RKKY method. Because of the use of full single-particle Green's functions, this method contains more many-body interactions than the conventional RKKY. Therefore it gives reliable results even for larger $J_H$.\\
\indent With this formalism it was possible to calculate Curie temperatures, which are in very good agreement to experimental measurements in the total ferromagnetic doping range, if we use intraband Coulomb repulsion. We have shown that therefore the additional terms to the KLM are essential to achieve these results. We found a phase diagram with FM-PI transitions where a CMR effect occurs and the JT distortion behaves qualitatively like in the measurements. But we have seen in our model that the neglection of interband Hubbard correlations will lead to an incorrect behavior of the resistivity. With the introduction of such interactions we get qualitatively correct results. A disadvantage of our effective medium treatment of the Hubbard part is, that it has a too strong influence on the magnetic properties, especially $T_c$. Possibly, a treatment of the Hubbard term that has a better inclusion of many-body correlations could correct these discrepancies.\\
\indent Most of the research on manganites is done with the simplification $J_H\rightarrow\infty$, which means effectively the neglection of the minority spins. But as can be seen in Fig. \ref{dos}, there exists also for $T<T_c$ a finite occupation of the spin-down band. This comes from scattering processes of the spin-down with the spin-up electrons accompanied by magnon emission or absorption, which is a result of the quantum mechanical treatment. Thus the influence of the spin-down electrons can not be neglected, just by the assumption of infinite Hund's coupling.\\ 
\indent In principle our model contains all ingredients to describe the para-/ferromagnetic phase of the manganites properly and none of the parts seems to be negligible. 
\\
\indent We are grateful to G.G. Reddy for fruitful discussions during his stay in Berlin.

\begin{appendix}

\section{\label{app_V}Hybridization}

\begin{figure}[tb]
\includegraphics[width=.95\linewidth]{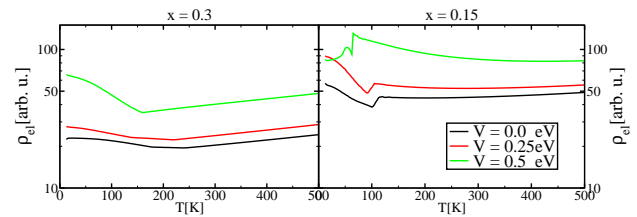}
\caption{\label{V_T}(color online) Resistivy curves for different hybridization values $V$ at two doping rates. Major discrepancies occur only if $V$ is much greater than the normal intraband hopping $t=0.16eV$. Parameters: $W=2eV,J_H=3eV,J_{AF}=0,g=0.5\sqrt{eV}$, interband repulsion (comp. Fig. \ref{rho_vgl})}
\end{figure}
\begin{figure}[tb]
\includegraphics[width=.95\linewidth]{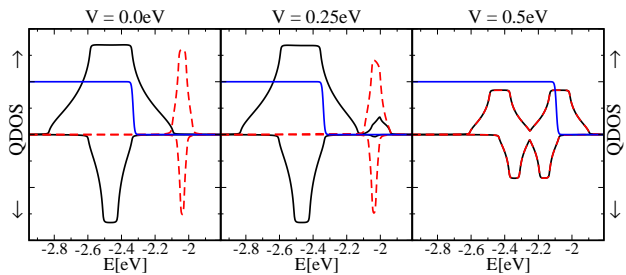}
\caption{\label{dos_V}(color online) Quasiparticle DOS for different hybridization values $V$. Lower JT-band is represented by the (black) solid line, upper by the (red) dashed line and the Fermi function by a (blue) solid one. Only the part at lower energies closed to the fermi edge is shown. For hybridizations in the near of the intraband hopping $t=0.16eV$ there are only small changes. Parameters like those in Fig. \ref{V_T} for $x=0.15$ and $T=50K$, interband Coulomb repulsion.}
\end{figure}

Hopping between different bands seems to be important in manganites. To recognize which effect such a hybridization could have in our model, we use an effective medium approach:
\begin{eqnarray}
\bar{\mathcal H} &=& \sksa (\eka + \Sigma_{\alpha\sigma}^{\text{ISA}}(E))\cpksa\cksa +\dots \nonumber\\ 
&&\dots+ \sksa V \cpksa c_{\fk\sigma-\alpha}
\end{eqnarray}
Here the first part is the original Hamiltonian (\ref{model}), represented by the approximative ISA self-energy (\ref{self}) containing the correlation effects. Due to the adding of $\Sigma_{\alpha\sigma}^{\text{ISA}}(E)$ the electrons can be treated like free electrons in an effective medium.  The second one describes the interband hopping with the hybridization $V$. After solving the according EOM we can investigate the possible changings due to this new term.\\
\indent There are no big changes of the resistivity curves for hybridization values at the order of magnitude of the intraband hopping ($t\approx 0.2eV$), as can be seen in Fig. \ref{V_T}. Only if $V$ exceeds this range it can come to essential modifications. This can be also shown at the density of states (Fig. \ref{dos_V}). The formation of a hybridization gap occurs only for larger $V$. For $V\lesssim t$ the QDOS is almost unchanged.\\
\indent Actually those bands we have investigated do not come from pure $d_{3z^2-r^2}$ or $d_{x^2-y^2}$ $e_g$- orbitals, but from the states shown in (\ref{states}). That means the resulting bands do not belong to one of those $e_g$-orbitals, but already contain some mixing.  

\section{Direct calculation of $T_c$}

By setting $\mean{S_z}\to 0^+$ in equation (\ref{sz_callen}) one can get an explicit formula for the Curie temperature
\begin{equation}
k_B T_c=\frac{2}{3}\hbar S(S+1) \left(\frac{1}{N}\sum_{\fq}\frac{1}{\left(\hat J(0)-\hat J(\fq)\right)_{T_c}}\right)^{-1}\label{tc}
\end{equation}
This formula (\ref{tc}) is exact, but it can cause problems if there are some temperature dependent variables, beside those of the KLM, which can influence the magnetization. In our case this can be the JT splitting at special parameter constellations. Thus (\ref{tc}) has to be handled carefully.

\section{Resistivity}

To calculate the electrical conductivity tensor, we can use the Kubo formula \cite{kubo} and get a current-current correlation function
\begin{equation}
\bar \sigma^{\alpha\beta}(E) = V\hspace{-.38cm}\int\limits^{(k_B T)^{-1}}_0 \hspace{-.38cm}d\lambda\int\limits_0^{\infty}\mean{j^{\beta}(0)j^{\alpha}(t+i\lambda\hbar)}e^{\frac{i}{\hbar}(E+i0^+)t}\ dt \nonumber
\end{equation}
For the special case $\ek = \epsilon (-\fk)$ and $v_{\alpha}(\fk)=\frac{1}{\hbar}\partial_{k_{\alpha}}\ek=-v_{\alpha}(-\fk)$, which holds for the simple cubic structure, this can be simplified to a formula, which only contains one-particle Green's functions \cite{velicky,rho}. With the definition of a transport function, e.g. in $x$-direction,
\begin{equation}
\phi(x) = \frac{1}{V} \sum_{\fk}\left(\frac{\partial\ek}{\partial k_x}\right)^2\delta(x-\ek)
\end{equation}
we now get
\begin{equation}
\bar\sigma(T)^{xx} \sim \frac{1}{k_B T}\sum_{\sigma\alpha}\int\!\!\!\!\!\intmax dE\ dx\ \frac{\bigl(\rho_{\sigma\alpha}(E,x)\bigr)^2}{4\cosh ^2(\frac{E-\mu}{2k_B T})}\phi(x) \label{sigma_el}\ ,
\end{equation}
with the QDOS $\rho_{\sigma\alpha}(E,x) =-\frac{1}{\pi}\text{Im}G^{\alpha\sigma}_x(E)$ and the resistivity $\rho_{\text{el}}(T)= \bar\sigma^{-1}(T)$. Results with the same structure can be found in other work\cite{rho1,rho2,rho3}.

\end{appendix}


\end{document}